# The Carbon Emissions of Writing and Illustrating Are Lower for AI than for Humans

**Authors:** Bill Tomlinson[1,2]*, Rebecca W. Black[1], Donald J. Patterson[1,3], Andrew W. Torrance[4,5]

**Affiliations:**

[1] Department of Informatics, University of California, Irvine, Irvine, CA 92697 USA.

[2] School of Information Management, Victoria University of Wellington - Te Herenga Waka, Wellington, New Zealand.

[3] Department of Mathematics and Computer Science, Westmont College, Santa Barbara, CA 93108 USA.

[4] School of Law, University of Kansas, Lawrence, KS 66045 USA.

[5] Sloan School of Management, Massachusetts Institute of Technology, Cambridge, MA 02142 USA.

* Corresponding author. Email: wmt@uci.edu.

# Abstract

As AI systems proliferate, their greenhouse gas emissions are an increasingly important concern for human societies. We analyze the emissions of several AI systems (ChatGPT, BLOOM, DALL-E2, Midjourney) relative to those of humans completing the same tasks. We find that an AI writing a page of text emits 130 to 1500 times less CO2e than a human doing so. Similarly, an AI creating an image emits 310 to 2900 times less. Emissions analysis do not account for social impacts such as professional displacement, legality, and rebound effects. In addition, AI is not a substitute for all human tasks. Nevertheless, at present, the use of AI holds the potential to carry out several major activities at much lower emission levels than can humans.

# Introduction

Artificial intelligence (AI) has made rapid advancements in recent years, with applications in a wide range of domains such as healthcare (*1*), finance (*2*), transportation (*3*), and environmental conservation (*4*). However, as the uses of AI have become more prevalent, concerns have also been raised about AI's detrimental impact on the environment, in particular the energy consumption required to train and run AI models and accompanying greenhouse gas emissions (e.g., *5, 6*). For example, training the model for GPT-3, one of the most powerful systems currently in broad deployment, produces emissions equivalent to the lifetime impact of five cars (*7*).

Several of the skills that AI is being trained to execute---such as the ability to write or to create images---are activities that previously were almost exclusively the domain of humans. In this article, we analyze the environmental impact of several AI systems in relative terms, comparing their emissions to those of humans completing the same task. Specifically, we focus on the tasks of writing and illustration. By comparing the environmental impact of these tasks when completed by AI versus humans, we highlight the substitutability between humans and AI, and demonstrate that, while AI has substantial environmental costs, at present these costs are typically far lower than for a human completing the same task.

We recognize that these findings are not generalizable to all contexts. While AI use may be beneficial in some writing and illustration contexts, not all activities lend themselves to AI intervention. In fact, AI and humans cooperating on tasks may remain the best approach in many fields. In addition, these findings are based on the current state of AI and human activity; future changes in technology and society will likely change the environmental impact of both AI (*8, 9*) and that of humans (*10*). There are also other complicating factors that need to be considered, such as professional displacement, legal use of training materials, and rebound effects. Nevertheless, the findings presented here suggest that concerns about the emissions generated by AI systems should be tempered by recognition that, even relying on cautious assumptions, humans produce far more emissions when engaging in some of the same tasks. While AI is often portrayed as an environmental threat to humanity, in this respect, at least, it may offer us valuable assistance.

# Results

## Writing: AI vs. Human

### AI Writing

While it can be difficult to define the scope of the problem when calculating the emissions produced by an AI system (5), two major components of that impact are the training of the model (a one-time cost that is amortized across many individual queries) and the per-query emissions. To offer two data points on the environmental impact of training models, training GPT-3 (the system on which the popular ChatGPT chatbot is based (*11*)) produces approximately 552 metric tons $CO_2e$ (*8*). Training BLOOM, a model slightly larger and substantially more energy-efficient than GPT-3, produces 30 metric tons of $CO_2e$ (*8*).

In addition to the amortized emissions of training, responding to each prompt carries its own emission footprint as well. An online estimate for ChatGPT (albeit an informal one) estimates that it produces 0.382 grams $CO_2e$ per query (*12*), based on 3.82 metric tons $CO_2e$ per day divided by 10,000,000 queries per day. A deployment of BLOOM produced 1.5 grams per query (340kg $CO_2e$ divided by 230,768 queries) (*9*).

Estimating that ChatGPT may do a full re-training of the model once per month, and continuing with the estimate of 10,000,000 queries per day, the 552 metric tons divided by 300,000,000 queries equates to 1.84 grams $CO_2e$ per query. Together, the training and operation for ChatGPT sum to 2.2 grams $CO_2e$ per queries. For BLOOM, assuming a similar level of usage and frequency of retraining as for ChatGPT, the per-query impact of training is 0.10 gram $CO_2e$, and the per-query cost is 1.47 grams, summing to 1.6 grams per query. These figures suggest that the impact of an AI query, including both amortized training and the query itself, is on the order of a few grams $CO_2e$.

## Human Writing

An article in The Writer magazine suggests that Mark Twain's output of approximately 300 words per hour, is "about the average when examining the daily work of other writers" (*13*). The emission footprint of a US resident is approximately 15 metric tons $CO_2e$ per year (*14*), or approximately 1.7kg $CO_2e$ per hour. Therefore, assuming that a person's emissions while writing are in line with their overall annual impact, we propose that the carbon footprint for a US resident generating a page of text (250 words) is approximately 1400 grams $CO_2e$. For a resident of India, by comparison, the annual impact is 1.9 metric tons (*14*), which equates to approximately 180 grams $CO_2e$ per page. We use the US and India here as the countries with the highest and lowest per capita impact among large countries (over 300M population).

In addition to the footprint of the person writing, the footprint of a computer running for the length of time it takes a human to write a page, approximately 0.8 hours, is itself far more impactful than the AIs. Assuming an average of 75 Watts for a typical laptop computer (*15*), the laptop produces 27 grams of $CO_2e$ (*16*). (We note that green energy providers may reduce the amount of $CO_2e$ resulting from this amount of computing, and that the EPA's Greenhouse Gas Equivalencies Calculator we used for this conversion masks a great deal of complexity on this topic. Nevertheless, for the purpose of comparison to humans, we assume that the EPA calculator captures the relationship adequately.) A desktop computer uses 200 Watts, producing 72 grams $CO_2e$ in the same amount of time.

Comparison

Figure 1 compares several variations of authorship: BLOOM is 1500 times less impactful, per page of text produced, than a US resident writing, and 190 times less impactful than a resident of India writing. ChatGPT is 1100 times less impactful than a US resident writing, and 130 times less impactful than a resident of India writing. Assuming the quality of writing produced by AI is sufficient for whatever task may be at hand, AI produces less $CO_2e$ per page than a human author.

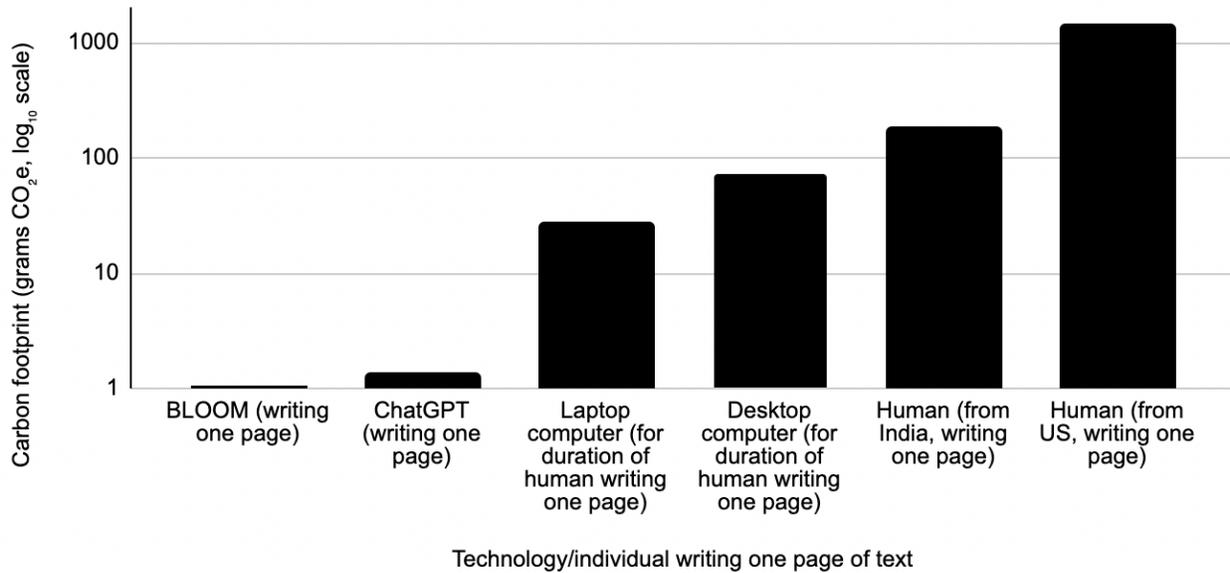

Figure 1: This figure compares the $CO_2e$ emissions of AI and humans engaged in the task of writing one page of text. AI writing (BLOOM or ChatGPT) produces 130 to 1400 times less $CO_2e$ per page than a human author. AI also produces substantially less $CO_2e$ than the computer usage to support humans doing that writing.

Authorship does not exist in a vacuum, and any accounting for the return on energy expenditure is confounded by the impact to the rest of the network in which it is embedded. For example, successful AI deployments may beget more costly models in the future, more frequent prompts by users, or more costly training schedules. On the other hand, human authorship may implicitly be training for other kinds of productive human work that would be lost in the face of the proliferation of AI writing. The freed human time may also incur new unexpected environmental costs.

# Illustration: AI vs. Human

Two prominent AI image generation engines are DALL-E2 and Midjourney. DALL-E2 is based on an underlying GPT-3 engine (similar to ChatGPT above), and Midjourney is based on a system called Stable Diffusion.

## AI Illustrator

Given the shared reliance on GPT-3, we estimate that DALL-E2's footprint is similar to the footprint of ChatGPT calculated above: 2.2 grams $CO_2$e per query. To estimate the impact of Midjourney, we take a different approach, based on statements made by Midjourney's CEO David Holz. Holz stated, with regard to Midjourney's computer usage, that "[e]very image is taking petaops … So 1000s of trillions of operations. I don't know exactly whether it's five or 10 or 50. But it's 1000s of trillions of operations to make an image… [W]ithout a doubt, there has never been a service before where a regular person is using this much compute" (*17*).

AI data centers, such as Google's Compute Engine (*18*), often run on Nvidia A100 GPUs (*19*). These GPUs can process 1.25 petaoperations per second while using 400 Watts of electricity (*19*). In the largest-emissions scenario (from Holz's comments), generating an image requires 50 petaoperations; therefore, the AI would need to run on that device for 40 seconds. This work would require 4.5Wh to process, or 1.9 grams $CO_2$e (*16*).

## Human Illustrator

There is a wide range of time that it may take for a human illustrator to produce an illustration. To arrive at an estimate for how long it takes, on average, we combined the average cost for an illustration project ($200 (*20*)), and the average hourly rate of pay for an illustrator ($62.50/hour (*20*)). Based on these figures, we propose that 3.2 hours per illustration is a viable estimate for a professional illustrator producing a commercial piece of work based on a provided specification, across a wide range of styles and formats. Since the environmental footprint for a US resident is approximately 15 metric tons $CO_2e$ per year (*14*), we calculate that the carbon footprint for a US-based illustrator is approximately 5500 grams $CO_2e$ per image. For a resident of India, by comparison, the impact would be approximately 690 grams $CO_2e$ per image (*14*).

The carbon footprint for a laptop operating for the duration of a human illustrator creating an image (3.2 hours) is 100 grams $CO_2e$. The footprint of that duration for a desktop computer is 280 grams $CO_2e$.

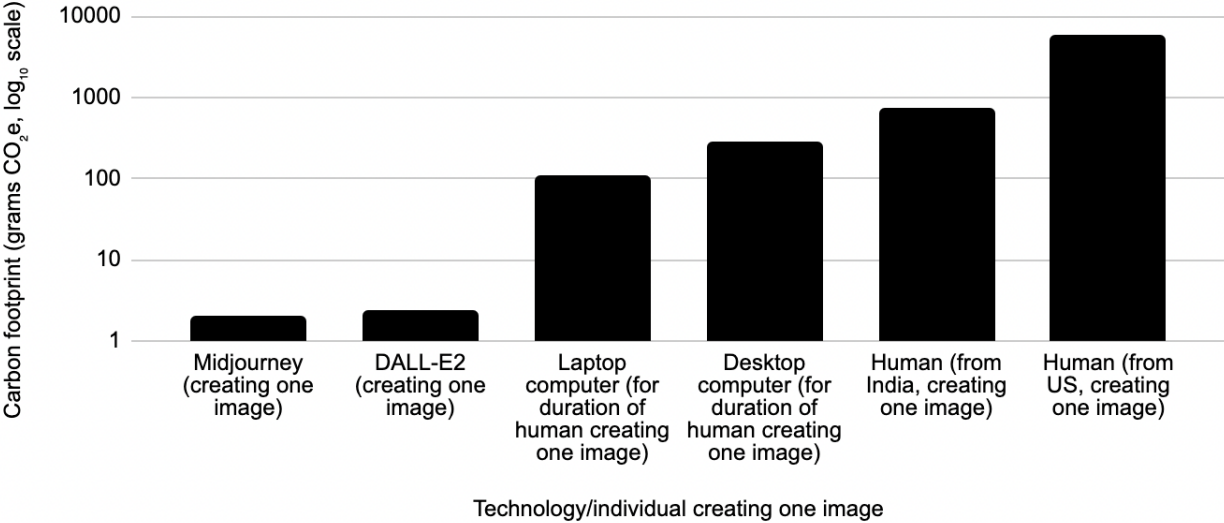

Figure 2: This figure compares the $CO_2e$ emissions of AI and humans engaged in the task of creating one image. AI image creation produces 310-2900 times less $CO_2e$ per image than human creators. AI produces many times less $CO_2e$ than computer usage to support humans making images.

## Comparison

Figure 2 shows that DALL-E2 emits approximately 2500 times less $CO_2e$ than a US-based artist, and approximately 310 times less than an India-based artist. Midjourney emits approximately 2900 times less $CO_2e$ than a US artist, and 370 times less than one based in India. Here, as with the writing analysis above, both laptop and desktop usage while supporting a human drawing an image would themselves be many times more impactful than the AI systems as well.

# Discussion

The findings above demonstrate that the environmental footprint of AI completing two major tasks is substantially lower than that of humans completing those same tasks.

The results for each specific task reflect an array of assumptions about the nature of these tasks and the people and AIs engaged in such tasks. For example, writing an in-depth, heavily-referenced, original article on a niche scientific topic is currently beyond the capabilities of an AI, and therefore is a context where human effort is more efficient than AI effort (since the AI cost for such a task is effectively infinite, at present). In the domain of illustration, drawing a stick figure is likely faster for a human than an AI at present (and therefore may have lower emissions due to dramatically lower speed), whereas the reverse is true for a complex illustration such as one resembling an oil painting. Nevertheless, despite these specific regions of the task-space where humans have lower emissions, the data presented in this paper suggest that, overall, the use of AI can significantly reduce the carbon footprint of certain tasks when compared to equivalent human activity.

These findings are also based on the current state of AI and human activity; future changes in technology and society will undoubtedly change the relative environmental impact of AI as well. For example, as evidenced by the order-of-magnitude difference in emissions from training GPT-3 (*8*) vs. training BLOOM (*9*) despite similar sized training data, algorithmic advances may profoundly reduce the footprint of AI systems, as has already been hypothesized (*8*). Alternatively, advances may improve the performance of AI, but at the cost of dramatic increases in energy use and accompanying emissions. For example, the possibility of ubiquitous personalization of AI content, in which all media consumed by everyone on earth---every book, every movie, every game, every educational worksheet---has been precisely tailored to that individual's unique and evolving preferences, paves the way for vastly greater emissions footprints for future AI systems. Whether the net effect of increasingly efficient algorithms and larger training sets and deployment contexts will cause total energy use to increase or decrease over time remains to be seen.

Similarly, societal changes regarding the footprint of various human societies may also change the ratio between AI and human activity. For example, the per capita impact of a human in the US has been mostly falling since it peaked in the 1970s (*14*), and the per capita impact of a human in India has been rising almost continuously since the 1940s (*14*) (although the impact of a resident of India is still only one seventh the impact of a US resident). These trends may continue, or may be altered by social and/or technical changes. Either way, they are highly likely to affect the human side of the AI/human ratio of the environmental costs related to the activities addressed in this article.

Previous studies have compared AI to other technological systems. Such comparisons obscure the role that AI is positioned to take in society, as AI transitions from digital tools of limited utility to more complex instruments with high generative capacity. AI is poised to take over roles once thought to be solely the domain of humans---those requiring creativity and the ability to integrate across multiple intellectual domains to synthesize concepts from each. This study provides new insights on the relative environmental footprint of AI and humans, and it highlights the importance of considering the impact of AI relative to a human when evaluating its overall impact on the environment.

While the environmental footprint of AI may be lower than that of humans for certain tasks, there are other important factors that may influence AI's overall impact on the world. For example, as AI technology becomes more advanced, it is likely that such technologies will displace human workers in certain industries. And, if the past is any indicator, professional displacement may lead to job losses and reduced income. The displacement of jobs by technology has been amply studied, e.g., (*21*), as has displacement by AI in particular (*22*). Job displacement is deeply problematic not only to those displaced, but to society at large, as it can disrupt the economic and social stability of entire geographic regions.

On the other hand, the development of AI has the potential to create jobs as well. These jobs could be meaningful and well-compensated replacements for those AI displaces, or they could be demeaning and/or involve low pay. For example, OpenAI, the creators of ChatGPT, outsourced work to a Kenyan company where workers were employed to label specific instances of toxic online content, including content many would likely find disturbing or distasteful, described as

"text [that] appeared to have been pulled from the darkest recesses of the internet" (*23*). Analogous displacements of workers took place during the Industrial Revolution and with the various technological revolutions accompanying the rise of digital technologies. While these displacements necessarily cause changes in the job industry, historically such technological shifts have given rise to new forms of employment to replace those lost.

There are also legal issues that are pertinent to the use of preexisting text, images, or sounds as training sets for AI. The legality of using preexisting material is particularly salient when training sets include copyrighted material, because use of such material may infringe. Perhaps "fair learning" will one day be recognized as a type of fair use that involves the transformation of copyrighted materials for educational purposes. However, at present, it remains unpredictable how courts will decide such a dispute. There is a class action lawsuit against the AI company Midjourney currently pending on this topic (*24*) that may provide precedent in this legal domain. Were Midjourney to be held liable for impropriety in using copyright works owned by others, the generous statutory damages scheme available to the plaintiffs could be ruinous for that particular company, while, more generally, chilling or inhibiting innovation in AI. On the other hand, if AI use of copyrighted material as training sets is held to be permissible, this will likely have the effect, within the current patent system, of spurring advances in AI. Another outcome could be the rise of companies acquiring vast sets of training data. While these legal issues are not necessarily intractable, they nevertheless represent an important point of contention over the future of such AI systems.

Additionally, as AI technology becomes more efficient, it is possible that such efficiency will lead to an increase in the demand for AI-produced goods and services, which could lead to further increases in resource use and pollution via rebound effects (*25*). The broadening of use cases for AI, and the proliferation of ways that AI could impact each use case (e.g., ubiquitous personalization of content) could lead to potentially far greater demand for energy than occurs at present.  As such, while the impact of AI is currently far less than humans in the tasks described above, it is important to maintain vigilance in this domain to avoid runaway resource use.  At the same time, it is possible that advances in the efficiency and specificity of AI could further decrease its environmental impact compared to human impacts from equivalent activities.  Such an increasing environmental advantage could argue in favor of accelerating applications of AI. In either scenario, vigilance and adaptation are vital.  And, whether the footprint of AI goes up or down, we support the call for disclosure of energy consumption to whatever degree possible across AI use cases (*26*).

Despite these current and potential future forms of societal transformation and harm, profound benefits to society could accrue through the use of AI.  Such systems could enable the development of new approaches to sustainable futures (*26*); they could lead to benefits in medicine (*27*); and they could improve human educational systems (*28*). We argue that these and other benefits of AI offset the potential harms such systems may entail.  And most relevant to the findings of this paper, AI can potentially do so with substantially lower carbon emissions.

We argue that the most beneficial and efficient use of both AI and human labor is via collaboration between the two types of entity, taking advantage of their respective strengths.  For example, in this article, we began with a draft written by an AI to bootstrap the effort, but the authors have edited it so thoroughly that the AI text is unrecognizable. (We acknowledge this use of AI for two reasons; first, it is required by the Nature submission guidelines and second, and perhaps more importantly, starting with AI was a more energy efficient way to achieve a high quality final product.)  Similarly, a human illustrator may choose to work with an AI in the early stages of an interaction with a client, to give them a sense of the broad range of possibilities available to them, and then complete a human-created illustration for the client only at the last stage.  Such a hybrid approach could allow for more rapid and more efficient coalescing of understanding between the client and the human illustrator, while also producing a final product that has the excellence and polish of a human-produced piece of work. (For example, unlike many AI-produced images, the human hands won't be uncannily misrepresented (*29*).) Hopefully such collaborative processes may address a range of concerns about AI-generated content (*30*).

In sum, due to its substantially lower impact than humans at at least two important tasks, AI can play an important role in various sectors of society without, at present, running afoul of problematic carbon emissions.  While the carbon footprint of AI is nontrivial, the footprint of humans doing the same work is far greater, and should not be discounted in the assessment of AI.

## Methods

We conducted numerical analyses based on previously published data for various aspects of the environmental impacts of both modern AI systems, humans in various locations around the world, and other components that may be involved in the production of text and images. These data sources were obtained from existing studies and databases on the environmental impact of AI and human activities. All figures and calculations are available below, or online here:

https://tinyurl.com/AICarbonEmissions

We used ChatGPT (Jan 9 version and Jan 30 version) as writing support in this article. We have run the text through the TurnItIn plagiarism detection software to ensure that ChatGPT did not inadvertently commit plagiarism or violate copyright.

# Acknowledgments

This material is based upon work supported by the National Science Foundation under Grant No. DUE-2121572.

# Supplementary Material

table S-1

| Data | Value | Units | Source |
| --- | --- | --- | --- |
| Training footprint, BLOOM | 30 | metric tons CO2e | https://arxiv.org/abs/2211.02001 |
| Usage footprint across 230,768 BLOOM queries | 340 | kg CO2e across 230,768 queries | https://arxiv.org/abs/2211.02001 |
| Usage footprint per query, BLOOM | 1.47 | grams CO2e | Derived from above and converted from kg to grams |

| Training footprint, ChatGPT | 552 | metric tons CO2e | https://arxiv.org/abs/2204.05149 |
|---|---|---|---|
| Usage footprint per day, ChatGPT | 3.82 | metric tons CO2e (across 10,000,000 queries) | https://medium.com/@chrispointon/the-carbon-footprint-of-chatgpt-e1bc14e4cc2a |
| Prompts per day, ChatGPT | 10,000,000 | prompts | https://medium.com/@chrispointon/the-carbon-footprint-of-chatgpt-e1bc14e4cc2a |
| Usage footprint per query, ChatGPT | 0.382 | grams CO2e | Derived from above and converted from metric tons to grams |
| Prompts per month | 300,000,000 | prompts | Derived from above |
| Amortized training of ChatGPT per query | 1.84 | grams/query | Derived from above |
| Amortized training of BLOOM per query | 0.10 | grams/query | Derived from above |
| Total footprint + amortized training of ChatGPT query | 2.22 | grams/query | Derived from above |
| Total footprint + amortized training of BLOOM query | 1.57 | grams/query | Derived from above |
| Watts for laptop | 75 | Watts | https://www.energuide.be/en/questions-answers/how-much-power-does-a-computer-use-and-how-much-co2-does-that-represent/54/ |
| Watts for desktop | 200 | Watts | https://www.energuide.be/en/questions-answers/how-much-power-does-a-computer-use-and-how-much-co2-does-that-represent/54/ |
| Writing (words/hour) | 300 | words/hour | https://www.writermag.com/writing-inspiration/the-writing-life/many-words-one-write-per-day/ |
| Writing (words/page) | 250 | words/page | Common knowledge, or https://wordcounter.net/words-per-page |
| Writing speed (hours/page) | 0.8333333333 | hours/page | Derived from above |
| US resident annual footprint | 15 | metric tons | https://ourworldindata.org/co2-and-other-greenhouse-gas-emissions |
| India resident annual footprint | 1.9 | metric tons | https://ourworldindata.org/co2-and-other-greenhouse-gas-emissions |
| Words per response (ChatGPT) | 412.8 | words | Average of several ChatGPT requests the research team posted (438+524+439+425+436+419+409+357+356+325)/10 |
| Footprint of laptop per hour | 32.4 | grams CO2e | https://www.epa.gov/energy/greenhouse-gas-equivalencies-calculator#results |
| Footprint of desktop per hour | 86.5 | grams CO2e | https://www.epa.gov/energy/greenhouse-gas-equivalencies-calculator#results |
| | | | |
| (Midjourney) Petaoperations per second for Nvidia A100 | 1.248 | petaoperations/second | https://www.nvidia.com/en-us/data-center/a100/ |
| (Midjourney) Electricity for Nvidia A100 | 400 | Watts | https://www.nvidia.com/en-us/data-center/a100/ |
| (Midjourney) Petaoperations per image | 50 | petaoperations/image | Derived from above |
| (Midjourney) Seconds per image | 40.06410256 | seconds/image | Derived from above |
| Wh per image | 4.451566952 | Wh/image | Derived from above |
| Cost of illustration project by human illustrator | 200 | USD | https://www.thumbtack.com/p/illustration-rates |
| Cost per hour for human illustrator | 62.5 | USD | https://www.thumbtack.com/p/illustration-rates |
| Hours per image for human illustrator | 3.2 | hours/image | Derived from above |
| Footprint of US resident per hour | 1712.328767 | grams CO2e | Derived from above and converted from metric tons to grams |
| Footprint of India resident per hour | 216.8949772 | grams CO2e | Derived from above and converted from metric tons to grams |
| | | | |
| **Writing results** | | | |
| Total footprint (including amortized training) of ChatGPT per page | 1.345687984 | grams CO2e/page | |
| Total footprint (including amortized training) of BLOOM per page | 0.9528471361 | grams CO2e/page | |
| Footprint of laptop per page | 27 | grams CO2e/page | |
| Footprint of desktop per page | 72.08333333 | grams CO2e/page | |
| Footprint of human per page (US) | 1426.940639 | grams CO2e/page | |
| Footprint of human per page (India) | 180.7458143 | grams CO2e/page | |
| Laptop vs. ChatGPT | 20.06408641 | | |
| Laptop vs. BLOOM | 28.33612967 | | |
| Desktop vs. ChatGPT | 53.56615662 | | |
| Desktop vs. BLOOM | 75.65046963 | | |
| US human vs. ChatGPT | 1060.380011 | | |
| US human vs. BLOOM | 1497.554629 | | |
| India human vs. ChatGPT | 134.3148013 | | |
| India human vs. BLOOM | 189.690253 | | |
| | | | |
| **Illustration results** | | | |
| Total footprint (including amortized training) DALL-E2 per image | 2.22 | grams CO2e/image | |
| Total footprint Midjourney per image | 1.9 | grams CO2e/image | https://www.epa.gov/energy/greenhouse-gas-equivalencies-calculator#results |
| Footprint of laptop per image | 103.68 | grams CO2e/image | |
| Footprint of desktop per image | 276.8 | grams CO2e/image | |
| Footprint of human per image (US) | 5479.452055 | grams CO2e/image | |
| Footprint of human per image (India) | 694.0639269 | grams CO2e/image | |
| Laptop vs. DALL-E2 | 46.66066607 | | |
| Laptop vs. Midjourney | 54.56842105 | | |
| Desktop vs. DALL-E2 | 124.5724572 | | |
| Desktop vs. Midjourney | 145.6842105 | | |
| US human vs. DALL-E2 | 2466.000025 | | |
| US human vs. Midjourney | 2883.922134 | | |
| India human vs. DALL-E2 | 312.3600031 | | |
| India human vs. Midjourney | 365.2968037 | | |
| **Creator** | **Carbon footprint (grams CO2e)** | | |
| BLOOM (writing one page) | 0.9528471361 | | |
| ChatGPT (writing one page) | 1.345687984 | | |
| Laptop computer (for duration of human writing one page) | 27 | | |
| Desktop computer (for duration of human writing one page) | 72.08333333 | | |
| Human (from India, writing one page) | 180.7458143 | | |
| Human (from US, writing one page) | 1426.940639 | | |
| | | | |
| Creator | Carbon footprint (grams CO2e) | | |
| Midjourney (creating one image) | 1.9 | | |
| DALL-E2 (creating one image) | 2.22 | | |
| Laptop computer (for duration of human creating one image) | 103.68 | | |
| Desktop computer (for duration of human creating one image) | 276.8 | | |
| Human (from India, creating one image) | 694.0639269 | | |
| Human (from US, creating one image) | 5479.452055 | | |